\documentclass[%
   aip,
   amsmath,amssymb,
    reprint,%
]{revtex4-1}
\usepackage{dcolumn}
\usepackage{bm}
\usepackage[utf8]{inputenc}
\usepackage[T1]{fontenc}
\usepackage{mathptmx}
\usepackage{color} 
\usepackage{xcolor}
\usepackage{graphicx} 
\usepackage{epsfig}

\newcommand{\bee}{\begin{equation}}
\newcommand{\ene}{\end{equation}}
\newcommand{\beea}{\begin{eqnarray}}
\newcommand{\enea}{\end{eqnarray}}

\usepackage{hyperref}
\hypersetup{colorlinks,linkcolor={blue},citecolor={blue},urlcolor={red}}  

\begin{document}
\title{Effect of Transverse Beam Size on the Wakefields and Driver Beam Dynamics
in Electron Beam Driven Plasma Wakefield Acceleration}
\author{Ratan Kumar Bera}
\email{rataniitb@gmail.com}
  \affiliation{ 
  Institute for Plasma Research, Bhat, Gandhinagar-382428, Gujarat, India.
 }%
\author{Devshree Mandal}
\affiliation{ 
Institute for Plasma Research, Bhat, Gandhinagar-382428, Gujarat, India.%
}%
\author{Amita Das}
\affiliation{%
Physics Department,
Indian Institute of Technology Delhi, Hauz Khas, New Delhi-110016, India. 
}%

\author{Sudip Sengupta}
\affiliation{ 
Institute for Plasma Research, Bhat, Gandhinagar-382428, Gujarat, India.%
}%

\date{\today}

\begin{abstract}  
In this paper, wakefields driven by a relativistic electron beam in a cold homogeneous plasma is studied
using 2-D fluid simulation techniques. It has been shown that in the limit
when the transverse size of a rigid beam is greater than the longitudinal extension, 
the wake wave 
acquires purely an electrostatic form and the simulation results show a good agreement with the 1-D
results given by Ratan et al. {\it [Phys. Plasmas, 22, 073109 (2015)]}. 
In the other limit, when the transverse dimensions are
equal or smaller than the longitudinal extension, 
the wake waves are electromagnetic in nature.
Furthermore,
a linear theoretical analysis of 2-D wakefields 
for a rigid bi-parabolic beam has also been done and compared with the simulations.
It has also been shown that the transformer ratio which
a key parameter that measures the efficiency in the process of acceleration, becomes higher
for a 2-D system ({\it i.e.} for a beam having a smaller transverse extension compared to longitudinal
length) than the 1-D system 
(beam having larger transverse extension compared to longitudinal
length). Furthermore,
including the self-consistent evolution of the driver beam in the simulation,
we have seen that the beam propagating inside the plasma undergoes the transverse pinching 
which occurs much earlier than the longitudinal
modification. Due to the presence of transverse dimensions in the system
the 1-D
rigidity limit 
given by Tsiklauri et al. [{\it Phys. Plasmas, 25, 032114 (2018)}] 
gets modified.  We have 
also demonstrated the modified rigidity limit
for the driver beam in a 2-D beam-plasma system.
\end{abstract}

\maketitle 
\section{ Introduction}
During the past two decades of research, the plasma wakefield acceleration (PWFA) scheme
has unarguably made amazing progress because of its wide
applications ranging from medical, industry, to high energy physics \cite{Edda,jing,esarey,Miller}. 
The scheme uses extremely large electric field associated with the
plasma waves to accelerate the charge particles. 
Typically these waves, commonly known as wake wave or wakefield,
are created injecting an ultra-relativistic electron beam inside a plasma.
The generation of strong electric field in a plasma
using an ultra-relativistic electron beam
stems from the pioneering work by Chen, Huff and Dawson \cite{chen}. Basically when ultra-relativistic electron 
beam propagates through a plasma, it expels the plasma electrons due to space charge force. 
Ions do not respond because of their heavy mass. As the beam moves further inside the plasma, 
these repelled electrons then try to come back to their original position. But due to their inertia, they 
overshoot and hence an oscillation will be established at the back of the beam. As a consequence, a wake wave will 
be excited at the wake of the beam \cite{Malka,katsouleas} propagating with a phase 
velocity equal to the velocity of the beam.
These waves, commonly known as wake wave or wakefield, are 
nothing but the disturbances left behind by the electron beam during the propagation
inside the plasma. 
Hence if a charge particle or beam rides this wave at an appropriate phase, 
it can be accelerated to high energy using the electric field associated with the wave.
As a relativistic electron beam is required even in the first place 
to excite the wake wave in PWFA, 
this scheme is not suitable to design a ``free-standing'' tabletop accelerator.
This scheme rather offers a way to boost the energy of the existing linacs \cite{Miller}. 
Success of this scheme has been demonstrated in a number of experiments \cite{Hogan,xia,Blumenfeld,litos} 
by accelerating electrons
up to GeV energies.
A great progress has also been made in applying PWFA concept to the astrophysical plasmas 
\cite{tsiklauri_1,tsiklauri_2,tajima_2,chen_astro}.
Several extensive numerical and analytical works have also been done in this area of research.
Most of these 
works have been done in 1-D where the variations transverse 
to the beam's directions are ignored. \cite{rosenzweig,ratan,tsiklauri}. 
On the other hand, fully 2-D nonlinear theoretical works on PWFA have mostly 
been done ignoring
the self-consistent evolution of the driver beam and using special beam profiles
\cite{lu,Baturin}.
Including the self-consistent evolution of the beam,
a details characteristic study
emphasizing the effect of finite transverse beam size on the
wakefield and the driver beam dynamics
is still largely an unexplored area of research.
\par\vspace{\baselineskip} 
 
In this paper, we report a detail study for the relativistic electron beam driven wakefield
in a cold plasma using 2-D fluid simulation techniques. 
We have performed the simulations over a wide range of 
beam parameter ({\it i.e.} density, velocity, 
longitudinal length, and transverse length). Here
longitudinal length refers to the length of the beam along the beam direction (longitudinal direction)
and transverse length is the length of the beam
perpendicular to the beam direction (transverse direction).
The simulations have been performed using rigid as well as self-evolving beam.
Rigid beam defines such a beam which can penetrate a long distance inside the plasma without 
any significant deformation \cite{ratan}.
In this case, the self-consistent evolution of the beam can be ignored. The beam behaves as a rigid piston.
For self-evolving beam configuration, the evolution of the beam in the self-consistent fields
has been considered. In the simulation 
when the transverse dimensions of a rigid bi-Gaussian beam are greater than their longitudinal extension, 
we have observed that the wake wave excited by the beam 
acquires purely an electrostatic form. The results obtained from 2-D simulation show a good agreement 
with the 1-D results given by 
Ratan et al.\cite{ratan}. On the other hand, when the transverse dimensions of a rigid beam is
equal or smaller than their longitudinal extension, 
the excitations become electromagnetic in nature and 2-D effects play
an important part in the dynamics of beam-plasma system. 
Furthermore, a linear theoretical analysis of 2-D wakefields 
for a rigid bi-parabolic
beam has also been done and the analytical results are compared with the fluid simulations.
We have also calculated the transformer ratio
which is a key parameter to measure the efficiency of the acceleration in the simulation.
We have observed that the value of the transformer ratio
is higher for 2-D cases (i.e. for a beam having transverse extension equal or smaller than its longitudinal
length) than 1-D systems ({\it i.e} for a beam having transverse extension larger than the longitudinal
length). It is to be noted that the transformer ratio for 2-D cases is more or less independent of beam density.
Including the self-consistent evolution of the driver beam in the simulations (fluid as well as Particle-In-Cell (PIC)),
we have also studied the rigidity of the driver beam. 
In this paper, PIC simulations are mainly used to support and validate the fluid results.
Using both these simulation techniques, it has been shown
that the relativistic driver beam propagating through a cold homogeneous plasma 
undergoes transverse pinching
that gives rise to the beam density
along beam direction. 
We have observed that the transverse pinching attributable to the transverse or focusing fields 
occurs much earlier than longitudinal 
modification during the propagation of a ultra-relativistic beam in plasma. 
It is also seen that
the transverse pinching depends on the beam velocity and occurs later for high beam velocity.
Due to the pinching in the transverse direction,
we have shown that the rigidity limit in terms of beam velocity ($v_{b}^r=0.9999 c$) 
given by Tsiklauri et al. \cite{tsiklauri}
given for 1-D cases gets modified. Rigidity limit $v_{b}^r$ defines such a beam velocity above which
the beam can be considered to be rigid for hundreds of plasma periods.
We have also shown the modified rigidity limit ($v_{b}^r=0.999999 c$) for a driver beam in 2-D system.

\par\vspace{\baselineskip}
The paper has been organized as following.
In Section-\ref{equations}, we present the basic equations governing the excitation of
relativistic electron beam driven wakefield in 2-D. We have discussed our numerical techniques used
for this study in section-\ref{simulation}. Our numerical observations and the detail discussion of
the obtained results has been covered in section-\ref{results}. 
We have summarized our studies in section-\ref{summary}.

\par\vspace{\baselineskip}
\section{Governing Equations}
\label{equations}
The basic equations governing the excitation of relativistic electron beam driven wakefield 
in a cold plasma in 2-D are the relativistic fluid-Maxwell equations. These equations contain 
the continuity and momentum equations for electron beam and plasma electrons.
Equations for plasma ions are ignored because of their heavy mass.
Maxwell's equations have been used to calculate fields in the system.
Therefore, the basic normalized equations for 2-D beam-plasma system can be written as,    
\begin{equation}
  \frac{\partial n}{\partial t}+\vec{\nabla}.(n \vec{v})=0	\label{pl_cont}
 \end{equation}
 \begin{equation}
  \frac{\partial \vec{p} }{\partial t}+(\vec{v}.\vec{\nabla}) \vec{p}=-\vec{E} -(\vec{v}\times \vec{B}) 
  \label{pl_mom}
 \end{equation}

  \begin{equation}
  \frac{\partial n_b}{\partial t}+\vec{\nabla}.(n_b \vec{v_b})=0	\label{beam_cont}
 \end{equation}
 \begin{equation}
   \frac{\partial \vec{p_b} }{\partial t}+(\vec{v_b}.\vec{\nabla}) \vec{p_b}=-\vec{E} -(\vec{v_b}\times \vec{B})  
   \label{beam_mom}   
 \end{equation}

  \begin{equation}
  \frac{\partial \vec{E} }{\partial t}= (n\vec{v} +n_b\vec{v_b})+ (\vec{\nabla} \times \vec{B})   \label{delte}
 \end{equation}

  \begin{equation}
  \frac{\partial \vec{B} }{\partial t}=-(\vec{\nabla} \times \vec{E})   \label{deltb}
 \end{equation}

 \begin{equation}
\vec{\nabla}.\vec{E}=(1-n-n_b) \label{pois}
 \end{equation}
\begin{equation}
\vec{\nabla}.\vec{B}=0 \label{divb}
 \end{equation}

where $\vec{p}=\gamma \vec{v}$ and  $\vec{p_b}=\gamma_b \vec{v_b}$  are the momentum 
of plasma electron and beam electron of velocity 
$\vec{v}$ and $\vec{v_b}$ respectively. Here, $\gamma=\left(1-v^2\right)^{-1/2}$ and
$\gamma_b=\left(1-v_b^2\right)^{-1/2}$ are the relativistic factors associated with
plasma electron of density $n$ and beam electron of density $n_b$ respectively.  
Other terms, $\vec{E}$ and $\vec{B}$ represents electric and magnetic field respectively. 
In the present study, we consider the 2-D slab geometry in ($x$, $z$)-plane where
the beam is moving along $z$-direction. 
We have used $\vec{\nabla}=  \hat{x}\frac{\partial}{\partial x} + \hat{z}\frac{\partial}{\partial z}$,
allowing the variation in both the longitudinal direction ($\hat{z}$) and transverse ($\hat{x}$). 
The above equations
use the following normalization factors.
$t \rightarrow \omega_{pe}t$, $(x,z) \rightarrow \frac{\omega_{pe}}{c}(x,z)$, 
$\vec{E} \rightarrow \frac{e\vec{E}}{m_e c\omega_{pe}}$, $\vec{v}\rightarrow \frac{\vec{v}}{c}$,
$\vec{v_b}\rightarrow \frac{\vec{v_b}}{c}$
$\vec{p} \rightarrow \frac{\vec{p}}{m_e c}$,
$\vec{p_b} \rightarrow \frac{\vec{p_b}}{m_e c}$, $n\rightarrow \frac{n}{n_0}$,
$n_i\rightarrow \frac{n_i}{n_0}$ and $n_b\rightarrow \frac{n_b}{n_0}$.
Equations (\ref{pl_cont}-\ref{divb}) are the main key equations required to examine 
the excitation of 2-D relativistic 
electron beam driven wakefield excitation in a cold plasma.
The exact analytical solution of these set of equations are formidable.
Therefore, we have solved these above equations numerically using fluid simulation techniques. 
For some specific cases, we have also employed 
fully electromagnetic particle based
Particle-In-Cell (PIC) simulation to compare the results obtained from the fluid simulation.
In next section, we have discussed both the fluid and PIC simulation techniques used to 
study the relativistic electron beam driven wakefield in a cold plasma.
\section{Simulation techniques}
\label{simulation}
In this section, we present numerical techniques used to study the relativistic 
electron beam driven wakefield excitation in a cold plasma. We have used both fluid simulation techniques
based on fluid description
of the $e^{-}$ beam-plasma medium as well as fully kinetic particle based 
Particle-In-Cell (PIC) simulation techniques. Below we briefly describe both the simulation techniques.

\subsection{ Fluid simulation of the relativistic electron beam driven wakefield}
\label{fluidsimulation}
We have developed a 2-D electromagnetic fluid code coupling iteratively a
set of subroutines based on flux-corrected
transport (LCPFCT) \cite{boris} scheme, Finite-difference time-domain (FDTD) method \cite{fdtd},
and Successive-Over-Relaxation (SOR) method \cite{sor}.
The basic principle of LCPFCT scheme
is based on the generalization of two-step Lax-Wendroff method \cite{numr}. 
LCPFCT method basically solves the generalized continuity type equations.
We have used this scheme to solve the continuity and momentum equations (\ref{pl_cont}-\ref{beam_mom}) 
for plasma electrons and electron beam.
FDTD method based on Yee algorithm has also been implemented 
to solve equations (\ref{delte}) and (\ref{deltb}).
The relaxation method has been used to solve equations (\ref{pois}) and (\ref{divb}).
In particular, equations (\ref{pois}) and (\ref{divb}) are solved to verify the obtained values of
electric and magnetic field from the simulations. 
Using this code, 
we have solved the equations (\ref{pl_cont}-\ref{divb}) numerically
with absorbing boundary conditions in both $x$ and $z-$ directions for all plasma parameters.
The time step ($\Delta t$) in the simulations has been chosen using the CFL condition \cite{boris}
$\Delta t= C_n \Delta S/u_{max}$;
where $C_n$, $\Delta S$, and $u_{max}$ are the Courant number, minimum grid size, 
and maximum fluid velocity respectively. In the simulations, we have used $C_n =0.2$, $\Delta S=\Delta z=\Delta x=0.05$ 
as the grid size $\Delta z$ and $\Delta x$
are same in both the directions, and $u_{max}=1$ as the maximum speed is equal to the speed of light.
In the simulation, the driver 
beam is initially kept just inside the plasma with the equilibrium values of plasma profiles at $t=0$. As time goes, 
the beam then propagates from one end (left) of the simulation window to its other end (right).  
As the beam passes the plasma, the wake wave is excited. We note
the profile of density and velocity for plasma electrons and beam, and components of electric field and magnetic field, 
with time.

\subsection{Particle-In-Cell (PIC) simulation of the relativistic electron beam driven wakefield}
For some special cases of studies (subsection \ref{self}),
we have used fully kinetic particle based Particle-In-Cell (PIC) simulation
techniques.
We have adopted fully explicit
electromagnetic massively parallel particle based PIC code ``OSIRIS 4.0'' \cite{Hemker,Fonseca, osiris} 
to study the propagation of electron beam 
in a plasma. 
For the same, a 2-D slab geometry has been chosen in $x-z$ plane;
where simulation box of $50L \times 100 L$  is considered,
where $ L(=c/\omega_{pe})$ is skin depth of the plasma. 
In this box, a bi-Gaussian beam propagating inside a homogeneous plasma 
along $+\hat{z}$ direction with the initial velocity of
$v_b$ has been taken. In the simulation, equilibrium
plasma density is considered as $n_0 = 10^{21}  cm^{-3} $. 
Plasma consists of electrons and ions (immobile) with a very 
small thermal velocity $ v_{th} = 0.0000442c$.
The simulation box has been divided into $ 5000 \times 2500 $ cells 
with $ 2 \times 2$ particles per cell and 
box has absorbing boundary conditions in both the directions for particle and fields.

\section{Results and discussion}
\label{results}
In this section, we present the key results obtained from simulations as well as from analytical analysis.
We have mostly used fluid simulation techniques for our study.
PIC simulations are used only for section (\ref{self}) to emphasize and validate
the fluid results for the self-consistent dynamics of the driver beam.
The fluid simulations have been performed over a wide range of beam parameters
for both the rigid beam and self-evolving beam.
In earlier 1-D works, it has been shown that the beam can be considered to be rigid if 
the velocity of the beam $v_b \rightarrow 1$  \cite{ratan,tsiklauri}.
In this limit, the momentum equation (\ref{beam_mom}) for the beam can
be ignored and the dynamics of a rigid beam can be depicted only by the equation (\ref{beam_cont}) with constant $v_b$.
To implement the rigid beam in the simulation, we have 
solved equations (\ref{pl_cont}-\ref{divb}) except the equation (\ref{beam_mom}) using $v_b=0.99999999$.
We have used rigid beam approximation for the first set of simulations
to illustrate the effect of finite transverse 
beam size on the excitation.
However, the rigidity of the driver beam is a central question of plasma wakefield acceleration scheme.
For an efficient acceleration process, the driver beam creating wakefield is
required to be rigid for a long distance. But in a real physical system, 
the driver beam eventually evolves in its self-consistent fields. 
In the second set of simulations, we have numerically 
solved the full set of equations (\ref{pl_cont}-\ref{divb}) for different beam velocity to study 
the rigidity limit for the driver beam in 2-D. Below we present these simulation results in detail.

\subsection{Effect of transverse beam dimensions on the excitation}
\label{beameffect}
\begin{figure*}
    \includegraphics[width=0.9\textwidth]{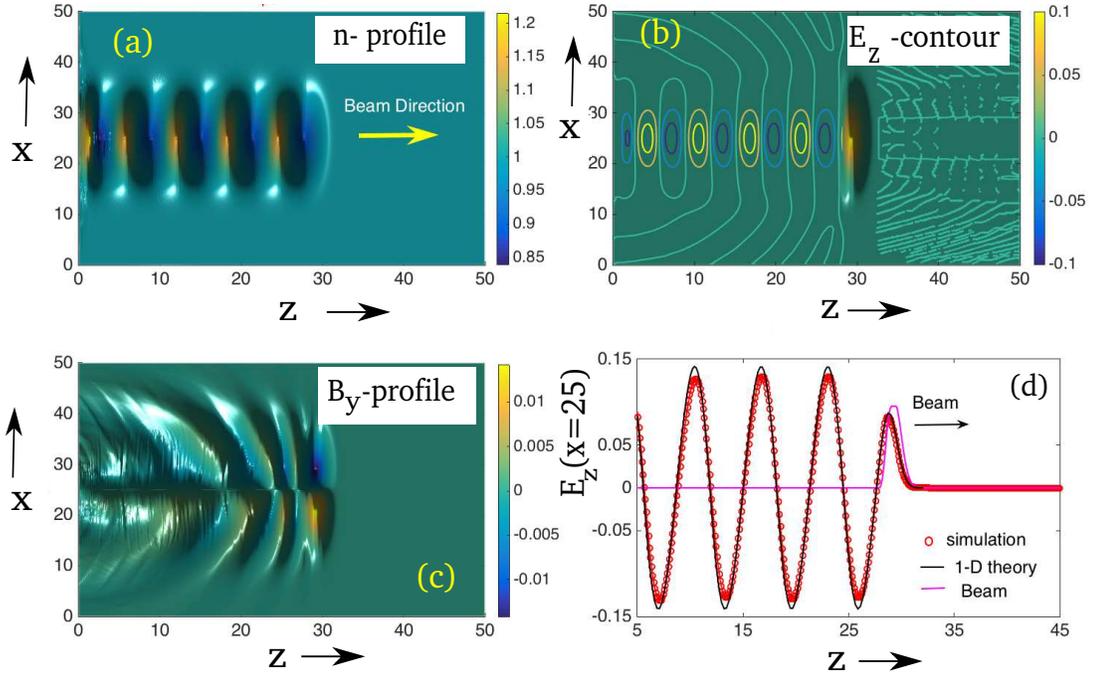}
    \caption{Plot of (a) normalized plasma electron density ($n$), (b) longitudinal electric field ($E_z$), 
    (c) $y-$ component of magnetic field ($B_y$)
    (d) profile of $E_z$ at $x=l_x/2$ as a function of $z$ from simulation (red circle) and 1-D theory (black solid line) 
    at $\omega_{pe}t = 25$ for the normalized peak beam density ($n_b$)=0.1, $\sigma_z=0.5$ and $\sigma_x=\sqrt{15}$, 
    and beam velocity ($v_b$) =0.99999999.}
    \label{fig1}
\end{figure*}

\begin{figure*}
    \includegraphics[width=0.9\textwidth]{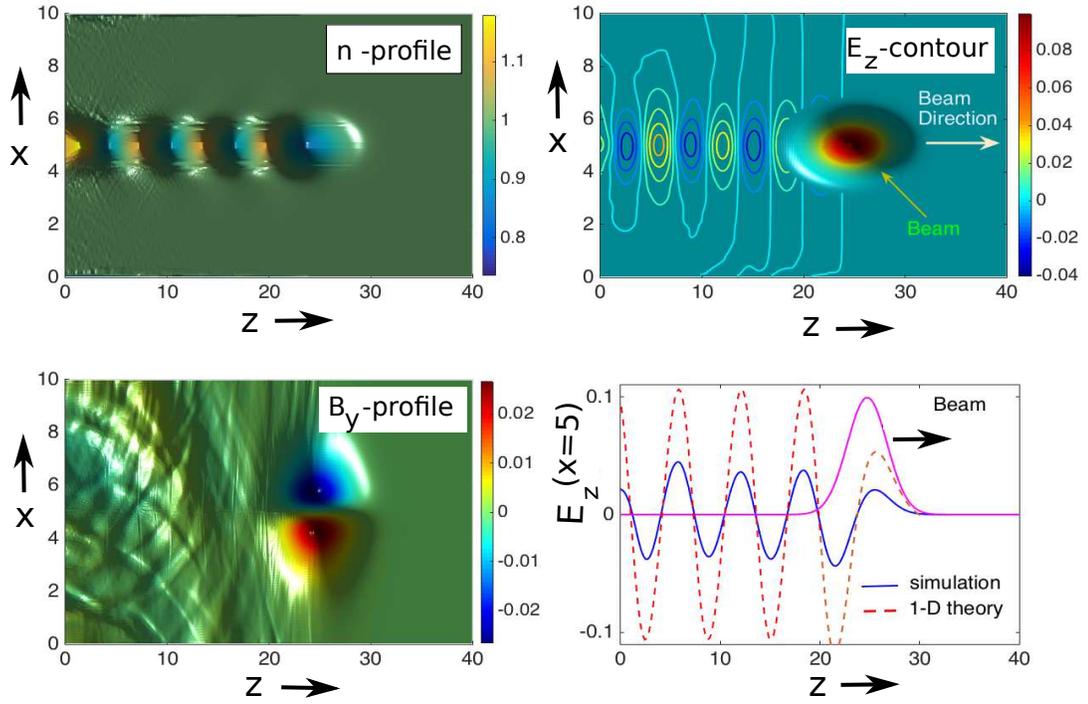}
    \caption{Plot of (a) normalized plasma electron density ($n$), (b) longitudinal electric field ($E_z$), 
    (c) $y-$ component of magnetic field ($B_y$),
    (d) profile of $E_z$ at $x=l_x/2$ as a function of $z$ from simulation (blue solid line) and 1-D theory (red dotted line) 
    at $\omega_{pe}t = 25$ for the normalized peak beam density ($n_b$)=0.1, $\sigma_z=\sqrt{5}$ and $\sigma_x=0.5$, 
    and beam velocity ($v_b$) =0.99999999.}
    \label{fig2}
\end{figure*}

First, we have studied the effect of transverse size of the driver beam
on the excited wakefield in a cold plasma.
We have run fluid simulations using a rigid, bi-Gaussian beam with different values of longitudinal and transverse beam
length. The beam velocity ($v_b$) for all the simulations presented in this subsection (\ref{beameffect}) 
is fixed and considered to be $0.99999999$ (or $\gamma_b=10000$).
The density profile of the beam has been chosen as,
\begin{equation}
 n_b =n_{b0} exp \left(-\frac{(x-\frac{l_x}{2})^2}{2 \sigma_x^2}\right)exp \left(-\frac{z^2}{2 \sigma_z^2}\right)
 \label{bigaussian}
\end{equation}

where $n_{b0}$ and $l_x$ define the peak density of the beam and
total length of simulation box along transverse direction $x$ respectively. Here 
$\sigma_z$ and $\sigma_x$ represent the 
extension or length of the beam along longitudinal and transverse direction respectively.
In Figs. (\ref{fig1}) and (\ref{fig2}) we show the fluid simulation results obtained 
for $n_{b0}=0.1$ at 
$\omega_{pe}t=25$. Fig. (\ref{fig1}) shows the plot of (a) the plasma density profile ($n$),
(b) longitudinal electric field ($E_z$), and (c) transverse magnetic field profile ($B_y$)
for $\sigma_z=0.5$ and $\sigma_x =\sqrt{15}$.
In Fig. (\ref{fig2}), we have plotted (a) the plasma density profile ($n$),
(b) longitudinal electric field ($E_z$), and (c) transverse magnetic field profile ($B_y$)
for $\sigma_z=\sqrt{5}$ and $\sigma_x =0.5$.
In the subplot (d) of both the Figs.(\ref{fig1}) and (\ref{fig2}), we have plotted the profile of
the longitudinal electric field ($E_z$)  at the mid of the beam ($x=l_x/2=25$) in transverse direction 
as a function of $z$  
obtained from the fluid simulations along with 
the 1-D analytical results. The 1-D results have been obtained from ref.\cite{rosenzweig}
using the Gaussian beam profile
$ n_b (z)=n_{b0} exp \left(-z^2/(2 \sigma_z^2)\right)$. We see that the 2-D simulation result 
shows a good agreement with the 1-D results for $\sigma_z=0.5$ and $\sigma_x =\sqrt{15}$ and deviates 
for $\sigma_z=\sqrt{5}$ and $\sigma_x =0.5$. This implies that 1-D treatment of beam-plasma system
is valid only for a beam having long transverse length compared to longitudinal extension.
In this case for $\sigma_z/\sigma_x <<1$ ,
the nature of the excitations acquires purely electrostatic form. However, the wakefields become
electromagnetic in nature for $\sigma_z/\sigma_x \geq 1$.
The 2-D effect plays important role and becomes unavoidable for a beam having transverse length smaller than its
longitudinal extension. 
Furthermore for the sake of completeness, an analytical treatment for 2-D linear wakefield has been also done. 
Below we discuss the analytical derivation of the linear wakefield for a rigid bi-parabolic beam.

\subsection{Analytical Description of 2-D wakefields}
\label{analytical}
Here we present a linear analytical calculation for the electron beam driven wakefield in a cold plasma in 2-D.
The analytical work has been done using
a bi-parabolic or an approximated bi-Gaussian beam having density profile, 
$n_b=n_{b0} (1-\frac{z^2}{b^2})(1-\frac{x^2}{a^2})$;
where $a$ and $b$ represents the extension of beam in longitudinal and transverse directions respectively.
Using the frame transformation ($\xi=z-v_bt$, $x$ ) and linearizing 
the equations (\ref{pl_cont}-\ref{deltb}), the evolution of plasma density inside the driver can be written as (see ref. 
\cite{chen_2d}),

\begin{equation}
\partial^2_\xi n_1(\xi,x) + n_1(\xi, x)=-n_b(\xi,x)=-n_{b0}g(\xi)f(x)
\label{n_1}
\end{equation}

The solution of the above equation is, 
\begin{equation}
n_1=n_{b0} f(x)G(\xi)
\end{equation}

Where $G(\xi)=\int_{0}^{\infty}  g(\xi')sin (\xi'-\xi) d\xi'$

In this frame, the equation for the evolution of longitudinal field is,
\begin{equation}
(\frac{d^2}{dx^2}-1)(A_{1z}-\phi_1)=-n_1
\end{equation}

where $\phi_1$ and $A_{1z}$ represents the scalar potential and $z$-component of vector potential respectively.
The solutions of the above equations inside the beam can be written as,
\begin{equation}
(A_{1z}-\phi_1)=-n_{b0} G(\xi) F(x)
\end{equation}
For the bi-parabolic beam profile, it is quite straightforward to find the value of $F(x)$ and $G(\xi)$.
Therefore, the solutions of the above equation can be written as,
\begin{equation}
\begin{split}
(A_{1z}-\phi_1)=-2n_{b0} [\frac{2}{a^2}+\frac{x^2}{a^2}-1]\\
\times[(1-\frac{(\xi+b)^2}{b^2})+\frac{2}{b} sin (\xi) + \frac{2}{b^2}(1-cos(\xi))]
\end{split}
\end{equation}

The expression of the  longitudinal electric field inside the beam is,
 \begin{equation}
 \begin{split}
E_z=\frac{\partial}{\partial \xi}(A_{1z-\phi_1})=-2n_{b0} [\frac{2}{a^2}+\frac{x^2}{a^2}-1]\\ 
 \times[(-\frac{2(\xi+b)}{b^2})
+\frac{2}{b} cos (\xi) + \frac{2}{b^2}sin(\xi)]\\
 \end{split}
 \end{equation}

At the wake of the beam ($n_b=0$), the solution of the equation (\ref{n_1}) can be written as, 
\begin{equation}
n_1=A(x) sin (\xi) +B(x) cos (\xi)
\end{equation}
Where $A(x)=n_{b0}F(x) (G(\xi_f)sin (\xi_f) + G'(\xi_f) cos (\xi_f))$ 
and $B(x)=n_{b0}F(x) G'(\xi_f)- \frac{A(y)}{F(x)tan (\xi_f)}$. 
Here $\xi_f$ represents the value of $\xi$ at the end of the beam. 
Thus the longitudinal electric field at the wake of the beam can be derived as,
 
\begin{equation}
E_z(\xi,x)=-2 A(x) cos (\xi) -2B(x) sin (\xi)
\label{2d_analytical}
\end{equation}

\begin{figure*}
    \includegraphics[width=0.9\textwidth]{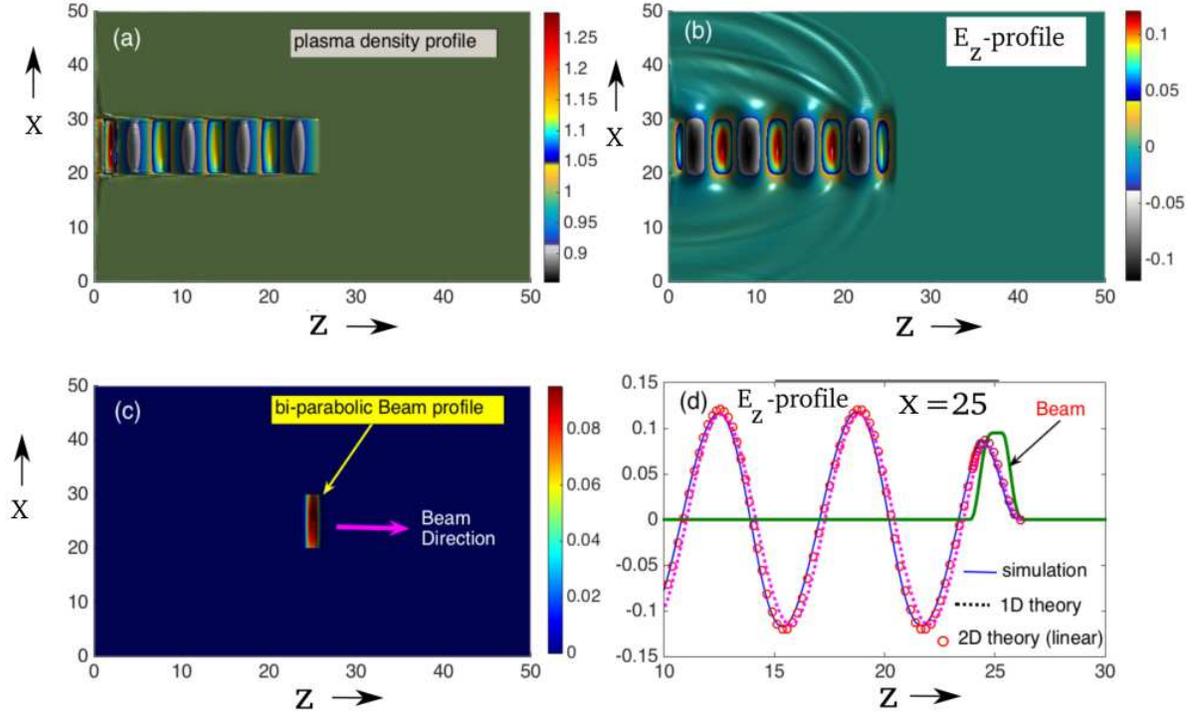}
   \caption{Plot of (a) normalized plasma electron density ($n$), (b) longitudinal electric field ($E_z$), 
    (c) beam ($n_b$),
    (d) profile of $E_z$ at $x=l_x/2$ as a function of $z$ from simulation (blue solid line), 2-D theory (red circle) and
    1-D theory (magenta dotted line) 
    at $\omega_{pe}t = 25$ for the normalized peak beam density ($n_b$)=0.1, $b=0.5$ and $a=\sqrt{15}$, 
    and beam velocity ($v_b$) =0.99999999.}
    \label{fig3}
\end{figure*}

\begin{figure*}
    \includegraphics[width=0.9\textwidth]{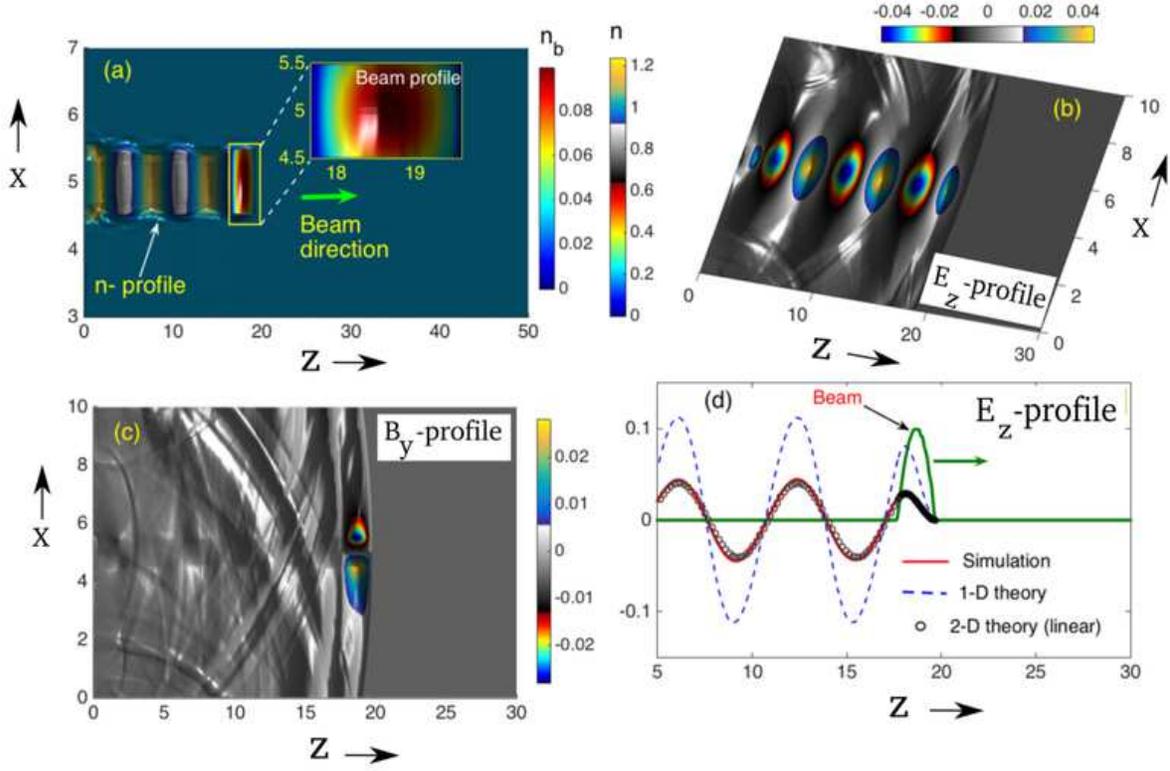}
     \caption{Plot of (a) normalized plasma electron density ($n$), (b) longitudinal electric field ($E_z$), 
    (c)$y-$ component of magnetic field ($B_y$),
    (d) profile of $E_z$ at $x=l_x/2$ as a function of $z$ from simulation (red solid line), 2-D theory (black circle) and
    1-D theory (blue dotted line) 
    at $\omega_{pe}t = 17$ for the normalized peak beam density ($n_b$)=0.1, $b=\sqrt{5}$ and $a=0.5$, 
    and beam velocity ($v_b$) =0.99999999.}
    \label{fig4}
\end{figure*}

Using bi-parabolic beam with peak beam density $n_{b0}=0.1$ and velocity $v_b=0.99999999$,
we have also performed fluid simulation for different values of longitudinal 
and transverse beam length. 
The simulation results are shown in (\ref{fig3}) and (\ref{fig4}). 
Fig. (\ref{fig3}) shows the profile of (a) plasma density ($n$), (b)longitudinal electric field ($E_z$), and
(c)beam density ($n_b$)
for 
$b=0.5$ and $a=\sqrt{15}$ at $\omega_{pe}t= 25$.
Fig. (\ref{fig4}) shows the profile of (a) plasma density ($n$), (b)longitudinal electric field ($E_z$), and
(c) transverse magnetic field profile ($B_y$)
for 
$b=\sqrt{5}$ and $a=0.5$ at $\omega_{pe}t=18$. For these two cases,
the ratio between longitudinal length to transverse length is
$b/a=0.129 <1$ and $l_s=b/a=4.4 >1$ respectively;.
In the last subplots of Figs.(\ref{fig3}(d)) and (\ref{fig4} (d)), 
we have plotted the profile of the longitudinal electric field at $x=l_x/2$ as a function of $z$
obtained from our simulation along with the profiles obtained from 2-D linear analytical results
obtained by solving equation (\ref{2d_analytical}) and corresponding 
1-D theoretical profile obtained from ref. \cite{rosenzweig}.
We have observed that our simulation result matches with the 2-D theoretical results for all the ratio of $b/a$. 
The profile of longitudinal electric field obtained from both the 2-D simulation and 
2-D theory deviates from 1-D theory results for $b/a=4.4 >1$. 
This implies that the extension of the beam decides the dimension of analysis for the beam-plasma system.
Earlier 1-D analysis of beam-plasma system 
\cite{ratan, tsiklauri, chen}
are only valid for a long transverse beam compared to its longitudinal extension. 

\par\vspace{\baselineskip}

\begin{figure*}
    \includegraphics[width=0.95\textwidth]{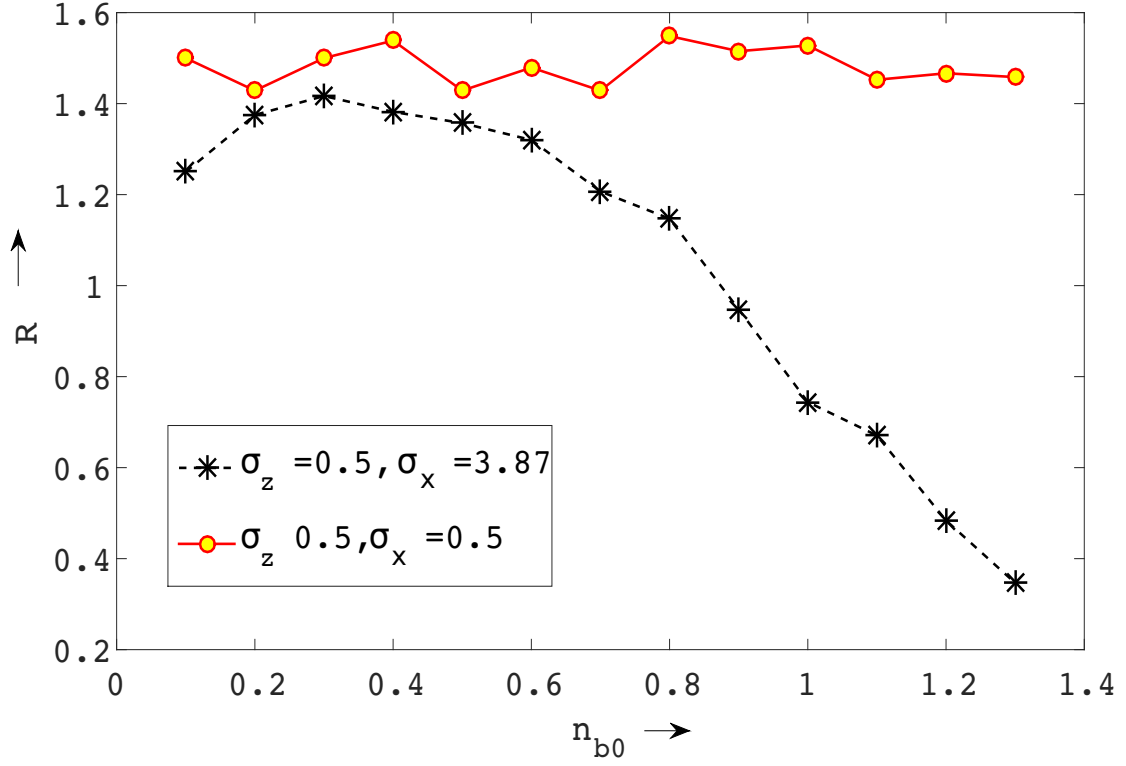}
    \caption{Plot of transformer ratio ($R$) as a function of peak beam density ($n_{b0}$) 
    for 1-D case ($\sigma_z=0.5$, $\sigma_x=\sqrt{15}$) 
    and 2-D case ($\sigma_z=0.5$, $\sigma_x=0.5$)}
        \label{fig5}

\end{figure*}

Furthermore, we have also calculated the transformer ratio using a bi-parabolic beam
for both 1-D (beam having $\sigma_z/\sigma_x <<1$) and 2-D
($\sigma_z/\sigma_x \geq 1$) system. In general, transformer ratio ($R$) is defined as, 
$R=E_{z}^-/E_{z}^+$; where $E_{z}^-$ and $E_{z}^+$ are the maximum accelerating 
outside the beam and maximum decelerating longitudinal electric field
field inside the beam respectively. We have calculated the values of 
$R$ from the simulation using a bi-parabolic beam of different length ratio ($\sigma_z/\sigma_x$).
In Fig. (\ref{fig5}), we have plotted the transformer ratio
as a function of different beam density ($n_b$) both for $\sigma_z/\sigma_x=0.5/3.87 <<1$ and 
$\sigma_z/\sigma_x=0.5/0.5=1$. 
Clearly, the excitation for $\sigma_z/\sigma_x <<1$ will be purely electrostatic or 1-D and the excitations will be 
electromagnetic in nature for $\sigma_z/\sigma_x=1$. 
We see that the transformer ratio obtained for 2-D cases is higher compared to that values obtained in 1-D cases.
It is also observed that the transformer ratio 
is nearly independent of the beam density for 2-D cases.  
Whereas the transformer ratio in 1-D cases monotonically decreases after a certain density (see also ref. \cite{ratan}).

\subsection{Self-Consistent evolution of the driver beam}
\label{self}
The results presented so far have been obtained using a rigid beam. This implies the 
charge density and energy of the beam remain
constant through out the propagation.
However in a realistic situation, 
the beam propagating inside the plasma must loose its energy and must be evolved in its self-consistent fields.
To avoid the beam deformation or self-consistent evolution 
of the beam inside the plasma, most of the analysis of PWFA
uses the beam velocity equal 
or very close to the speed of light . Including the 
self-consistent evolution of the beam,  several authors also have extensively studied the beam dynamics
in a plasma in 1-D and reported a critical beam velocity $v_b^r$ above which the beam can be considered to 
be rigid \cite{ratan, tsiklauri}. Recently
using 1-D PIC simulation, Tsiklauri et al. \cite{tsiklauri} established that the driver beam would behave rigid
when the velocity of the beam $v_b \geq 0.9999$ or
$\gamma_b \geq 70.7$. 

\par\vspace{\baselineskip}

\begin{figure*}
    \includegraphics[width=0.9\textwidth]{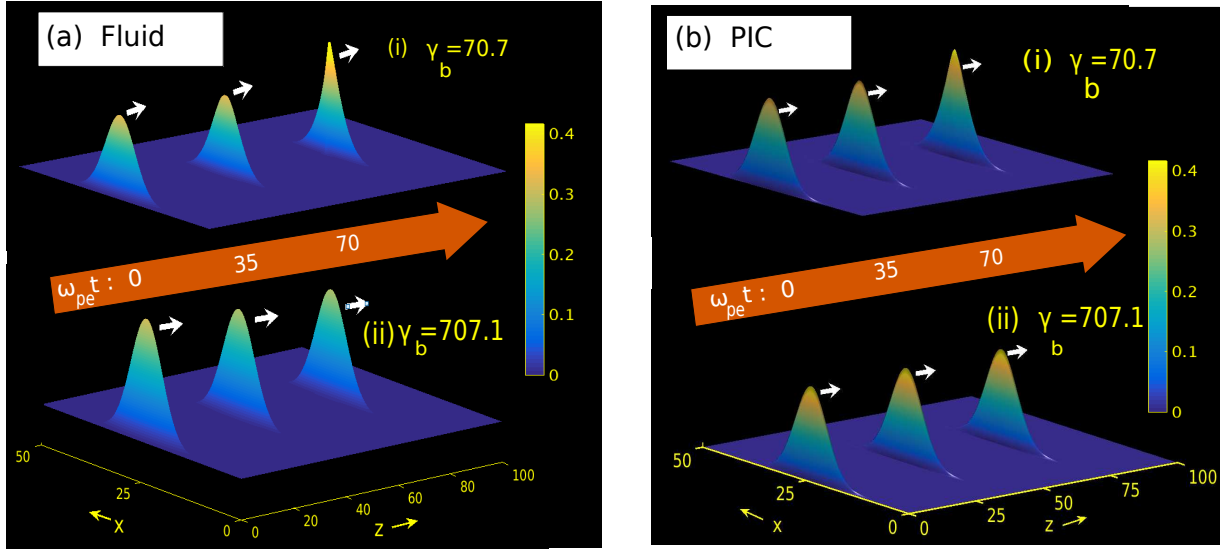}
    \caption{Plot of beam density profile ($n_b$) having $n_{b0}=0.3$, $\sigma_z=0.5$, $\sigma_x=\sqrt{15}$ 
    at different times for
    $\gamma_b=70.7$ and $\gamma_b =707.1$
     using (a) Fluid simulation and (b) PIC simulation. Subplots a(i) and b(i) represent the beam evolution  for
     $\gamma_b=70.7$ 
     in Fluid and PIC simulation respectively. Subplots a(ii) and b(ii) represents the beam evolution for
     $\gamma_b=707.1$ in Fluid and PIC simulation respectively.}
    \label{fig6}
\end{figure*}

\begin{figure*}
    \includegraphics[width=0.9\textwidth]{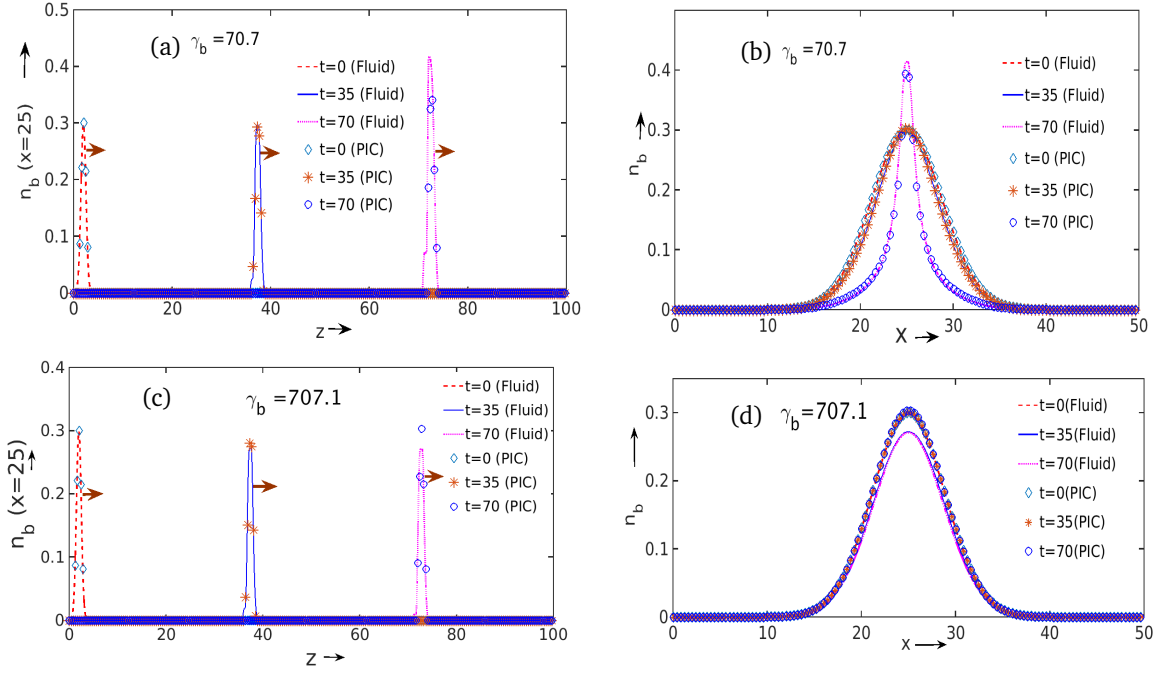}
    \caption{Plot of (a) the beam density profile ($n_b$) at $x=l_x/2=25$ as a function of $z$
    obtained from both Fluid and PIC simulation at different times for $\gamma_b=70.7$, 
    (b) the beam density profile ($n_b$) at $z_t$ (value of $z$ at the peak of the beam) as a function of $x$ obtained from both Fluid and PIC simulation 
    at different times for $\gamma_b=70.7$.
    (c)the beam density profile ($n_b$) obtained from both Fluid and PIC simulation 
    as a function of $z$ at different times for $\gamma_b=707.1$, 
    (b) the beam density profile ($n_b$) at $z_t$ (value of $z$ at the peak of the beam) 
    as a function of $x$ obtained from both Fluid and PIC simulation 
    at different times for $\gamma_b=707.1$.}
    \label{fig7}
\end{figure*}

In this subsection we present a detail investigation on the rigidity of the driver beam including
the self-consistent evolution in 2-D.
We have performed both fluid and PIC simulations to study the driver beam dynamics in a plasma.
In the fluid simulation, we have solved the 
full set of equations (\ref{pl_cont}-\ref{divb}) using a bi-Gaussian beam profile defined by 
equation (\ref{bigaussian}) for different beam velocity.
In the PIC simulations, we have also used same beam and plasma profiles and parameters.
The details of the PIC simulation techniques used here are described in the section (\ref{simulation}).
PIC simulations are mainly used to verify the fluid results as analytical solution
of the full set of equations (\ref{pl_cont}-\ref{divb}) is difficult.
The simulation results are presented in Figs. (\ref{fig6}) and (\ref{fig7}) for a bi-Gaussian beam
having $n_{b0}=0.3$, $\sigma_z=0.5$, and $\sigma_x=\sqrt{15}$. Therefore, we have performed simulations for 1-D cases.
as $\sigma_z/\sigma_x <<1$. In Fig.(\ref{fig6} (a))
we have plotted the driver
beam density profile at different times obtained from fluid simulation for beam velocities $v_b=0.9999$
or $\gamma_b=70.7$(top) and  
$v_b=0.999999$ or $v_b=707.1$ (bottom). In Fig.(\ref{fig6} (b))
we have shown the driver
beam density profile at different times obtained from PIC simulation for beam velocities $v_b=0.9999$
or $\gamma_b=70.7$ (top) and  
$v_b=0.999999$ or $v_b=707.1$ (bottom). 
It is evident from both these simulations that the beam profile gets modified
for $v_b=0.9999$ within hundred of plasma periods even for a long transverse beam. Whereas it can be considered to
be rigid
for $v_b=0.999999$ or $v_b=707.1$. We note that
these results presented here are in contrast with the results 
given by Tsiklauri et.al. \cite{tsiklauri}.
This is due to the presence of transverse dimension in the simulations.
Tsiklauri et al. using 1-D PIC simulation showed that 
the beam having velocity $v_b=0.9999$ must be rigid for hundreds of plasma periods inside a plasma. 
They have deliberately restricted the dynamics of the beam only in 
one dimension ignoring the variation in transverse direction.
Although the effective dimension of the problem is still 1-D (as
$\sigma_z/\sigma_x <<1$), but due to the presence of transverse variation in the simulation
the beam having finite transverse extension gets modified.
In a real physical system, we always have a beam of finite extensions.
Due to the finite transverse extension which can be much larger than longitudinal length,
the evolution of the beam acquires an additional modification by the transverse dynamics.
\par\vspace{\baselineskip}

To understand the beam evolution in detail, 
we have also plotted the beam profiles in Fig. (\ref{fig7}) in both longitudinal and transverse direction 
from both the fluid and PIC simulation. 
In Fig. (\ref{fig7}) we have plotted the beam profiles at different times for 
$\gamma_b=70.7$ and $\gamma_b=707.1$. 
In Fig. (\ref{fig7} (a)), we show  the beam density profile ($n_b$) at $x=l_x/2=25$ as a function of $z$
obtained from both Fluid and PIC simulation at different times for $\gamma_b=70.7$.
Fig. (\ref{fig7} (b)),
plots the beam density profile ($n_b$) at $z=z_t$ as a function of $x$ obtained from both Fluid and PIC simulation 
at different times for $\gamma_b=70.7$. 
Here $z_t$ defines the value of $z$ where the peak of the beam lies at different times.
In Fig. (\ref{fig7}) (c), we have plotted the beam density profile ($n_b$) at $x=l_x/2=25$ as a function of $z$
obtained from both Fluid and PIC simulation at different times for $\gamma_b=707.1$.
Fig. (\ref{fig7}) shows
the beam density profile ($n_b$) at $z=z_t$ as a function of $x$ obtained from both Fluid and PIC simulation 
at different times for $\gamma_b=707.1$. 
We see that the beam undergoes transverse pinching during the propagation inside a plasma. 
The transverse pinching is an well known phenomena in relativistic beam-plasma system \cite{Rhon}.
The transverse pinching occurs due to the finite size in the transverse direction. Because of finiteness
in the transverse direction,
the system possess radial wakefield or focusing fields that gradually pinches the beam.
As a consequence, the density of beam beam increases along longitudinal direction. It is seen that
the longitudinal length of the beam remains unchanged for both $v_b=0.9999$ and $v_b=0.999999$.
We also note that transverse pinching modifying the beam profile depends on the beam velocity.
For high beam velocity, a significant pinching occurs in later times.
We see that the effect of transverse pinching is negligibly small for hundreds of plasma periods for $v_b=0.999999$ and 
the beam is rigid.
The results presented in the subsection (\ref{beameffect}) using the rigid beam are consistent with these studies
as as they are obtained for
$v_b =0.99999999$ or $\gamma_b=10000$.
\section{Conclusion}
\label{summary}
As a summary, this paper reports 
a detail study of electron beam driven wakefield in a cold plasma in 2-D
using fluid as well as Particle-In-Cell(PIC) simulation techniques. It has been shown that in the limit 
when the transverse dimensions of a rigid bi-Gaussian beam are greater than their longitudinal extension, 
the wake wave excited by the beam 
acquires purely an electrostatic form and the simulation results show a good agreement with the 1-D results given by 
Ratan et al.\cite{ratan}. In the other limit, when the transverse dimensions of a rigid beam is
equal or smaller than their longitudinal extension, 
the wakefields become electromagnetic in nature.
Furthermore, a linear theoretical analysis of 2-D wakefields 
has been carried out for a rigid bi-parabolic
beam and the results have been verified with the simulations.
The transformer ratio which is a key parameter for measuring the efficiency of the acceleration in PWFA,
is found to be larger for 2-D cases than that obtained from 1-D systems. 
Including the self-consistent beam evolution in the simulation,
we have studied the driver beam dynamics in 2-D system. It has been shown that the rigidity limit of the driver beam 
obtained from purely 1-D simulations system gets modified due to the presence of transverse dimension.
The relativistic beam propagating inside the plasma gets pinched in the transverse dimension.
We have also obtained and demonstrated the modified rigidity limit for the driver beam in PWFA 
concept. 

\section{Acknowledgement}
The authors are grateful to the IPR computer center
where the simulation studies presented in this paper were
carried out.  A.D and D.M would like to acknowledge the OSIRIS Consortium,
consisting of UCLA and IST (Lisbon, Portugal) for providing access to the
OSIRIS 4.0 framework \cite{Fonseca,osiris} which is supported by NSF ACI-1339893.
A.D would also like to acknowledge the financial support of her J C Bose Fellowship grant
JCB/2017/000055 of DST for the work. 

%
\bibliographystyle{unsrt}

\begin{thebibliography}{10}
\bibitem{Edda}
Edda Gschwendtner and Patric Muggli, 
\newblock Plasma wakefield accelerators,
\newblock {\em Nature Reviews Physics}, 1, 246–248 (2019).

\bibitem{jing}
C. Jing,
\newblock  Dielectric wakefield accelerators,
\newblock {\em Rev. Accel. Sci.
Techol.}, 09, 127 (2016).


\bibitem{esarey}
E.~Esarey, C.~B. Schroeder, and W.~P. Leemans,
\newblock Physics of laser-driven plasma-based electron accelerators,
\newblock {\em Rev. Mod. Phys.}, 81, 1229(2009).

\bibitem{Miller}
Johanna L. Miller,
\newblock Plasma wakefield acceleration shows promise,
\newblock {\em Physics Today}, 68, 1, 11 (2015).



\bibitem{chen}
Pisin Chen, J.~M. Dawson, W.~Robert Huff and T.~Katsouleas,
\newblock Acceleration of electrons by the interaction of a bunched electron beam with a plasma,
\newblock {\em Physical review letters}, 54, 693(1985).

\bibitem{Malka}
Chan Joshi and victor Malka,
\newblock Focus on Laser- and Beam-Driven Plasma Accelerators,
\newblock {\em New J. Phys.}, 12, 045003(2010).



\bibitem{katsouleas}
T.~Katsouleas,
\newblock Physical mechanisms in the plasma wake-field accelerator,
\newblock {\em Phys. Rev. A}, 33, 2056(1986).



\bibitem{Hogan}
M. J. Hogan, C. D. Barnes, C. E. Clayton, F. J. Decker, 
S. Deng, P. Emma, C. Huang, R. H. Iverson, D. K. Johnson, C. Joshi, T. Katsouleas, 
P. Krejcik, W. Lu, K. A. Marsh, W. B. Mori, P. Muggli, C. L. O’Connell, E. Oz, R. H. Siemann, and D. Walz,
\newblock Multi-GeV Energy Gain in a Plasma-Wakefield Accelerator,
\newblock {\em Phys. Rev. Lett.} 95, 054802 (2005).

\bibitem{xia}
G. Xia
, D. Angal-Kalinin 
, J. Clarke 
, J. Smith 
, E. Cormier-Michel 
, J. Jones 
,
P.H. Williams 
, J.W. Mckenzie 
, B.L. Militsyn 
, K. Hanahoe 
, O. Mete
, A. Aimidula
,
C.P. Welsch,
\newblock A plasma wakefield acceleration experiment using CLARA beam,
\newblock {\em Nuclear Instruments and Methods in Physics Research Section A: Accelerators, 
Spectrometers, Detectors and Associated Equipment} 775, 168 (2015).



\bibitem{Blumenfeld}
Ian Blumenfeld, Christopher E. Clayton, Franz-Josef Decker, Mark J. Hogan, Chengkun Huang,
Rasmus Ischebeck, Richard Iverson, Chandrashekhar Joshi, Thomas Katsouleas, Neil Kirby,
Wei Lu, Kenneth A. Marsh, Warren B. Mori, Patric Muggli, Erdem Oz, Robert H. Siemann, Dieter Walz and Miaomiao Zhou,
\newblock Energy doubling of 42 GeV electrons in a metre-scale plasma wakefield accelerator,
\newblock {\em Nature}, 445, 741(2007).

\bibitem{litos}
M.~Litos, E.~Adli, W.~An, C.~I. Clarke, C.~E.Clayton, S.~Corde, 
J.~P. Delahaye,	R.~J. England, A.~S. Fisher, J.~Frederico, S.~Gessner,	
S.~Z. Green, M.~J. Hogan, C.~Joshi,	
W.~Lu,	K.~A. Marsh, W.~B. Mori,	
P.~Muggli, N.~Vafaei-Najafabadi,	
D.~Walz, G.~White, Z.~Wu, V.~Yakimenko and G.~Yocky,
\newblock High-efficiency acceleration of an electron beam in a plasma wakefield accelerator,
\newblock {\em Nature}, 515, 92(2014).


\bibitem{tsiklauri_1}
David Tsiklauri,
\newblock Electron plasma wake field acceleration in solar coronal and chromospheric plasmas,
\newblock {\em Phys. of Plasmas}, 24, 072902 (2017).

\bibitem{tsiklauri_2}
David Tsiklauri,
\newblock Collisionless, phase-mixed, dispersive, Gaussian Alfven pulse in transversely inhomogeneous plasma
,
\newblock {\em Phys. of Plasmas}, 23, 122906 (2016).

\bibitem{tajima_2}
T. Ebisuzaki and T. Tajima,
\newblock Astrophysical ZeV acceleration in the relativistic jet from an accreting supermassive blackhole
,
\newblock {\em Astroparticle Physics}, 56, 9-15 (April 2014).

\bibitem{chen_astro}
Pisin Chen, Toshiki Tajima, and Yoshiyuki Takahashi,
\newblock Plasma Wakefield Acceleration for Ultrahigh-Energy Cosmic Rays,
\newblock {\em Phys. Rev. Lett.}, 89, 161101 (2002).


\bibitem{rosenzweig}
J.~B. Rosenzweig,
\newblock Nonlinear Plasma Dynamics in the Plasma Wake-Field Accelerator,
\newblock {\em Physical Review Letters}, 58, 555(1987).

\bibitem{ratan}
Ratan Kumar Bera, Sudip Sengupta and Amita Das,
\newblock Fluid simulation of relativistic electron beam driven wakefield in a cold plasma,
\newblock {\em Phys. of Plasmas}, 22, 073109 (2015).


\bibitem{tsiklauri}
David Tsiklauri,
\newblock Differences in 1D electron plasma wake field acceleration in MeV versus GeV and linear versus blowout regimes
,
\newblock {\em Phys. of Plasmas}, 25, 032114 (2018).


\bibitem{lu}
W.~Lu, C.~Huang, M.~Zhou, M.~Tzoufras, F.~S. Tsung, W.~B. Mori,T.~Katsouleas,
\newblock A nonlinear theory for multidimensional relativistic plasma wave wakefields,
\newblock {\em Physics of Plasmas}, 13, 056709(2006).


\bibitem{Baturin}
S. S. Baturin, G. Andonian, and J.B. Rosenzweig,
\newblock Analytical treatment of the wakefields driven by transversely shaped beams
in a planar slow-wave structure,
\newblock {\em PHYS. REV. ACCEL. BEAMS}, 21, 121302 (2018).





\bibitem{boris}
Jay.~P. Boris, Alexandra M. Landsberg, Elaine S. Oran and John H. Gardner,
\newblock LCPFCT- Flux-corrected Transport Algorithm for Solving Generalized Continuity Equations,
\newblock {\em Naval Research laboratory, Washington}, NRL/MR/6410-93-7192, 1993.

\bibitem{fdtd}
U. S. Inan and R. A. Marshall,
\newblock Numerical Electromagnetics: The FDTD Method,
\newblock {\em Cambridge University Press, 2011}.

\bibitem{sor}
A. Hadjidimos,
\newblock Successive overrelaxation (SOR) and related methods,
\newblock {\em Journal of Computational and Applied Mathematics}, 123, 177-199 (2000)



\bibitem{numr}
W.~Press, R.~Assmann, A.~Teukolsky, W.~Vetterling and Brian P. Flannery,
\newblock Numerical Recipes: The Art of Scientific Computing.
\newblock {\em Cambridge University Press}, 1992.

\bibitem{Hemker}
R. G. Hemker,
\newblock Particle-In-Cell Modeling of Plasma-Based Accelerators in Two and Three Dimensions
\newblock {\em Thesis, University of California, Los Angeles}, 2000.
\newblock (\em Physics - Computational Physics, Physics - Plasma Physics)

\bibitem{Fonseca}
R. A. Fonseca,
L.O. Silva,
F.S. Tsung, V.K. Decyk,
W. Lu,
C. Ren,
W.B. Mori,
S. Deng,
S. Lee,
T. Katsouleas,
and J.C. Adam,
\newblock OSIRIS: A Three-Dimensional, Fully Relativistic Particle in Cell Code for Modeling Plasma Based Accelerators,
\newblock {\em Computational Science --- ICCS 2002: International Conference
Amsterdam, The Netherlands, April 21--24, 2002 Proceedings, Part III}.

\bibitem{osiris}
R. A. Fonseca, S. F. Martins, L. O. Silva, J. W. Tonge, F. S. Tsung, and W. B. Mori
\newblock One-to-one direct modeling of experiments and astrophysical scenarios: pushing
the envelope on kinetic plasma simulations
\newblock {\em Plasma Physics and Controlled Fusion}, 50, 12 (124034), 2008.


\bibitem{chen_2d}
Pisin Chen,
\newblock A possible final focusing mechanism for linear colliders,
\newblock {\em SLAC-PUB-3823 (Rev.)}, SLAC/AP-46, (A/AP), November (1985).


\bibitem{Rhon}
Rhon Keinigs, and Michael E. Jones,
\newblock Two-dimensional dynamics of the plasma wakefield accelerator
,
\newblock {\em Physics of Fluids}, 30, 252 (1987).

\end{thebibliography}

%
\newpage
%
\par\vspace{\baselineskip}

\end{document}